\documentclass[%
 reprint,
 amsmath,amssymb,
 aps,
]{revtex4-2}

\usepackage{graphicx}% Include figure files
\usepackage{subcaption}
\usepackage{dcolumn}% Align table columns on decimal point
\usepackage{bm}% bold math
\usepackage{comment}
\usepackage[capitalize]{cleveref}

\usepackage[usenames, dvipsnames]{xcolor}

\newcommand{\mnras}{Monthly Notices of the Royal Astronomical Society}
\newcommand{\jcap}{Journal of Cosmology and Astroparticle Physics}

\begin{document}

\preprint{APS/123-QED}

\title{Cosmological simulations of mixed ultralight dark matter}% Force line breaks with \\

\author{Alex Lagu\"e}%
\email{alague@sas.upenn.edu}
\affiliation{Department of Physics and Astronomy, University of Pennsylvania,
209 South 33rd Street, Philadelphia, PA 19104-6396, USA}
\affiliation{Dunlap Institute for Astronomy \& Astrophysics, University of Toronto,
50 St. George Street, Toronto, ON M5S 3H4, Canada}
\affiliation{Canadian Institute for Theoretical Astrophysics, University of Toronto,
60 St. George Street, Toronto, ON M5S 3H8, Canada}
\affiliation{David A. Dunlap Department of Astronomy \& Astrophysics, University of Toronto,
50 St. George Street, Toronto, ON M5S 3H4, Canada}%
 
\author{Bodo Schwabe}
%\email{bschwabe@unizar.es}
\affiliation{Departamento de F\'isica Te\'orica, Universidad de Zaragoza, 50009 Zaragoza}
\affiliation{Institut f\"ur Astrophysik,  Universit\"at  G\"ottingen, Germany}

\author{Ren\'ee Hlo\v zek}
%\email{bschwabe@unizar.es}
\affiliation{Dunlap Institute for Astronomy \& Astrophysics, University of Toronto,
50 St. George Street, Toronto, ON M5S 3H4, Canada}
\affiliation{David A. Dunlap Department of Astronomy \& Astrophysics, University of Toronto,
50 St. George Street, Toronto, ON M5S 3H4, Canada}%

\author{David J. E. Marsh}
\affiliation{Theoretical Particle Physics and Cosmology, King’s College London, Strand, London, WC2R 2LS, United Kingdom}

\author{Keir K. Rogers}
\affiliation{Dunlap Institute for Astronomy \& Astrophysics, University of Toronto,
50 St. George Street, Toronto, ON M5S 3H4, Canada}

\date{\today}% It is always \today, today,

\begin{abstract}
The era of precision cosmology allows us to test the composition of the dark matter. Mixed ultralight or fuzzy dark matter (FDM) is a cosmological model with dark matter composed of a combination of particles of mass $m\leq 10^{-20}$ eV, with an astrophysical de Broglie wavelength, and particles with a negligible wavelength sharing the properties of cold dark matter (CDM). In this work, we simulate cosmological volumes with a dark matter wave function for the ultralight component coupled gravitationally to CDM particles. We investigate the impact of a mixture of CDM and FDM in various proportions ($0\%,\;1\%,\;10\%,\;50\%,\;100\%$) and for ultralight particle masses ranging over five orders of magnitude ($2.5\times 10^{-25}\;\mathrm{eV}-2.5\times 10^{-21}\;\mathrm{eV}$). To track the evolution of density perturbations in the non-linear regime, we adapt the simulation code \textsc{AxioNyx} to solve the CDM dynamics coupled to a FDM wave function obeying the Schr\"odinger-Poisson equations. We obtain the non-linear power spectrum and study the impact of the wave effects on the growth of structure on different scales. We confirm that the steady-state solution of the Schr\"odinger-Poisson system holds at the center of halos in the presence of a CDM component when it composes 50\% or less of the dark matter but find no stable density core when the FDM accounts for 10\% or less of the dark matter. We implement a modified friends-of-friends halo finder and find good agreement between the observed halo abundance and the predictions from the adapted halo model \textsc{axionHMCode}. 

%\begin{description}
%\end{description}
\end{abstract}

\maketitle

%\tableofcontents

%% INTRO %%

\section{Introduction}
The cold dark matter (CDM) model explains the formation of cosmic structures exceptionally well on large scales. Observations suggest that dark matter may consist of multiple components, with varying properties affecting the growth of structure on different scales~\cite{Abdalla2022CosmologyIntertwined}. Among the dark matter candidates which cluster distinctly from  CDM are very light scalar bosons with negligible non-gravitational interactions known as fuzzy dark matter (FDM)~\cite{Hu2000FuzzyCold,Rogers2023Ultra-lightAxions}. Due to their low particle mass, FDM condensates are subject to coherence effects which impede clustering on scales below their de Broglie wavelength $\hbar/mv$ (for a particle of mass $m$ and velocity $v$), but behave as CDM on large scales. Ideal candidates for FDM from particle physics are ultralight axions which arise naturally in quantum chromodynamics~\cite{Preskill1983CosmologyOf,Abbott1983ACosmological,Dine1983TheNotSoHarmless,Berezhiani1992PrimordialBackground,Kim2016AnUltralight,Davoudiasl2017FuzzyDark} and in high energy physics extensions to the Standard Model such as string theory~\cite{Arvanitaki2010StringAxiverse,Mehta2021SuperradianceIn,Cicoli2022FuzzyDark,Gendler2023GlimmersFrom}. In this scenario, it is postulated that they form in a \textit{plenitude} of $\mathcal{O}(100)$ axion fields with logarithmically distributed masses~\cite{Arvanitaki2010StringAxiverse}. Other extensions of the Standard Model not based on string theory, such as the pi-axiverse~\cite{Alexander2023ThePiAxion}, can also lead to a set of ultralight bosons. In both cases, the lightest of these axions could have a mass lower than $10^{-19}$ eV, implying the existence of particles with de Broglie wavelengths on galactic scales. In the high energy physics models where they arise, the ultralight particles' relic density, which we will denote $\Omega_\mathrm{FDM}$, is not necessarily equal to the total dark matter density of the Universe. For instance, Ref.~\cite{Bachlechner2018Multiple-Axion} finds an axion with a mass of $2.5\times 10^{-22}$ eV and a cosmological density about one-tenth of the predicted total dark matter density from the $\Lambda$CDM model. Throughout this work, we make the simplifying assumption that only one of the particles produced in this scenario is ultralight while the others have negligible de Broglie wavelengths. In this case, we group their combined relic density into $\Omega_\mathrm{CDM}$ since they can be modeled as CDM by virtue of the Schr\"odinger-Vlasov correspondence~\cite{Mocz2018Schrodinger-Poisson}.

\begin{figure*}
    \centering
    \includegraphics[width=0.8\textwidth]{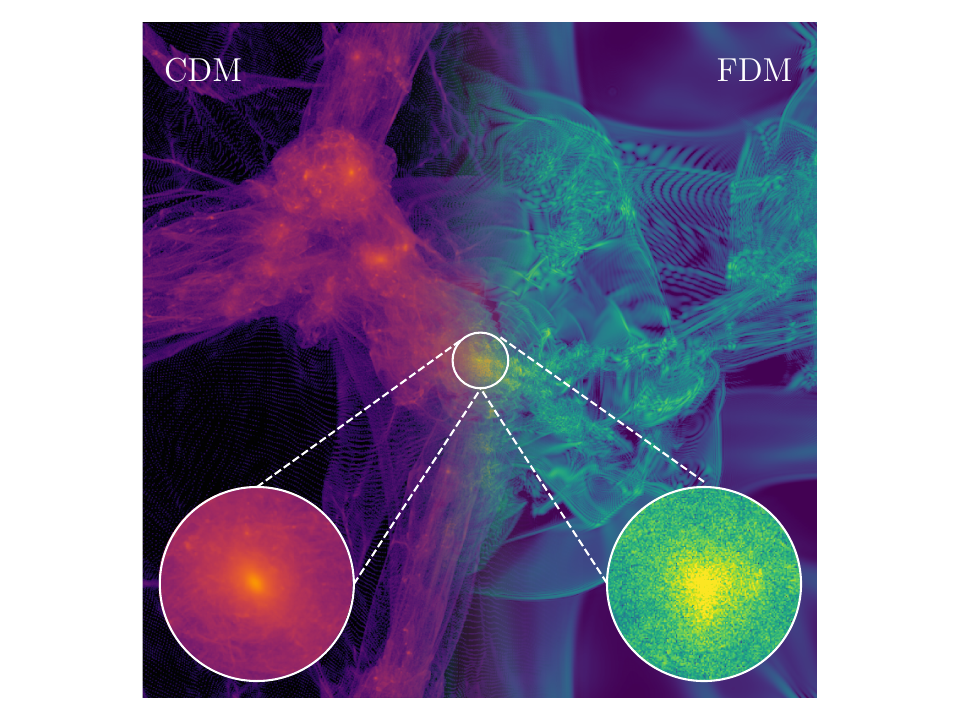}
    \caption{Side-by-side slice plot of CDM and FDM in the same 1 Mpc/$h$ box (comoving) at redshift $z=4$. The dark matter is composed of 10\% FDM, with the rest being CDM. The inner region illustrates the difference in clustering between the two dark matter species inside the central halo.}%\alex{confirm box size and adjust table}}
    \label{fig:half-half}
\end{figure*}
\begin{figure*}
    \centering
    \includegraphics[width=1.0\textwidth,]{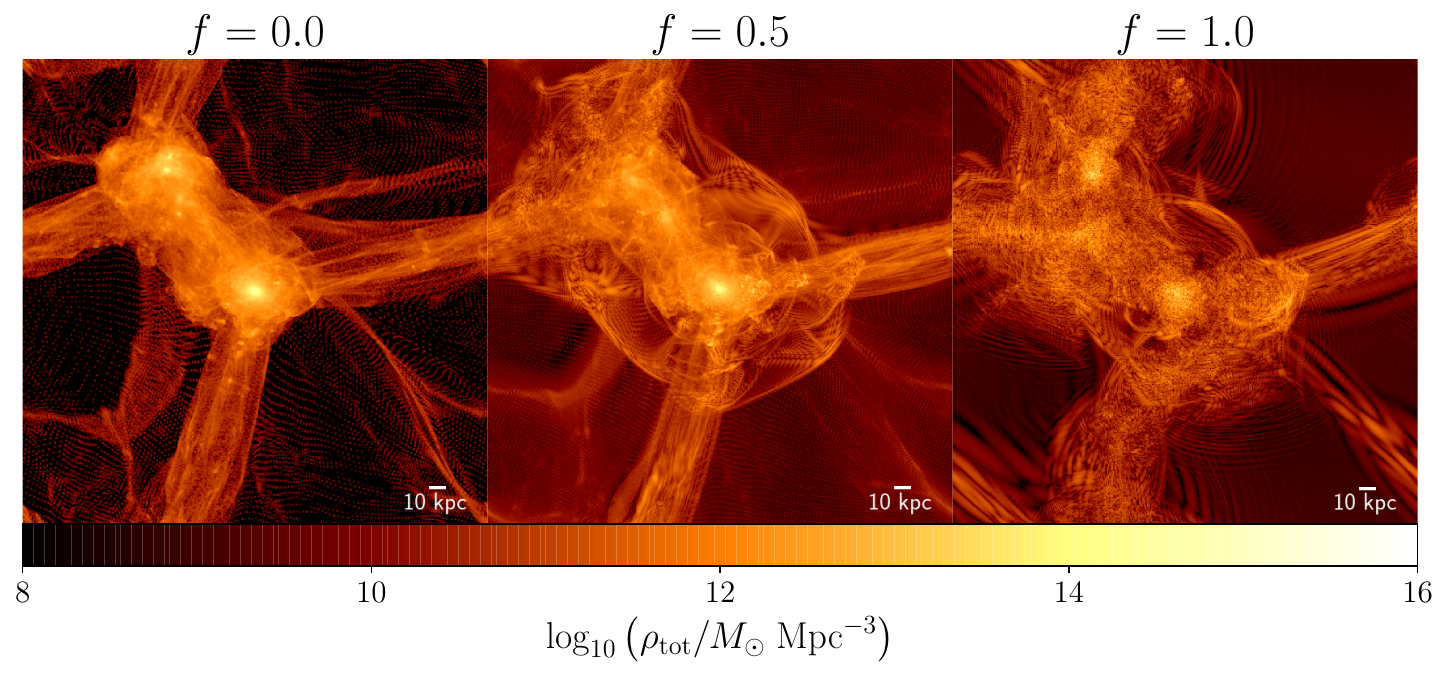}
    \caption{Density slices of 1 Mpc/$h$ simulation boxes (comoving) at redshift $z=4$. The three panels (starting from the left) illustrate the total (CDM plus FDM) density around the central halo for a simulation of pure CDM, an even mixture of CDM and FDM, and of pure FDM for a FDM mass of $2.5\times 10^{-22}$ eV.}
    \label{fig:density_slice}
\end{figure*}

With the total relic density of the non-ultralight particles captured by $\Omega_\mathrm{CDM}$, we define the FDM fraction as
\begin{align}
    f \equiv \frac{\Omega_\mathrm{FDM}}{\Omega_\mathrm{CDM}+\Omega_\mathrm{FDM}}.
\end{align}
The mixed cold and fuzzy dark matter model has been studied in numerical simulations for single halos~\cite{Schwabe2020SimulatingMixed,Gosenca2023MultifieldUltralight}, but cosmological $N$-body and hydrodynamical simulations (accounting for FDM dynamics) up to date have assumed $f=1$~\cite{Nori2018Ax-Gadget,Mocz2019FirstStar,Veltmaat2018FormationAnd,Hopkins2019AStable}. In this work, we investigate the full non-linear behaviour of FDM particles with cosmological initial conditions while foregoing this assumption. Previews of the simulation with multiple dark matter components are shown in Figs.~\ref{fig:half-half} and \ref{fig:density_slice}.

Constraints on the FDM particle mass and fraction have been reached with CMB~\cite{Hlozek2015AData,Hlozek2018UsingThe,Rogers2023Ultra-lightAxions}, galaxy clustering~\cite{Lague2022ConstrainingUltralight,Rogers2023Ultra-lightAxions}, galaxy weak lensing~\cite{Dentler2022FuzzyDark}, and Lyman-$\alpha$ forest data~\cite{Irsic2017FirstConstraints,Armengaud2017ConstrainingThe,Rogers2021StrongBound}. With the Lyman-$\alpha$ forest, accounting for the full evolution of the FDM wave function is crucial to arrive at unbiased constraints in the presence of baryons~\cite{Zhang2018TheImportance}. While constraints using the Lyman-$\alpha$ forest were obtained for mixed and pure FDM scenarios, only the case where $f=1$ has been verified with $N$-body simulations which included a non-linear treatment of the wave effects of FDM~\cite{Nori2019LymanAlpha}. Studies of the density profiles of ultra-faint dwarf galaxies indicate a preference for a FDM mass $m\sim 3.7-5.6\times 10^{-22}$ eV~\cite{Calabrese2016UltraLight} while dwarf spheroidal galaxies suggest $m \leq 1.1 \times 10^{-22} $ eV~\cite{Safarzadeh2020Ultra-Light}, and Lyman-$\alpha$ forest results give $m\geq 3.8\times 10^{-21}$ eV~\cite{Irsic2017FirstConstraints} and $m > 2\times 10^{-20}$ eV~\cite{Rogers2021StrongBound}, respectively. Furthermore, the suppression of smaller halos conflicts with the subhalo mass function for ultralight dark matter masses below $m\sim 2.1\times 10^{-21}$ eV~\cite{Schutz2020SubhaloMass}. It is also in tension with measurements of stellar streams in the Milky Way~\cite{Dalal2020DontCross}. However, these arguments are based on simulations of a single dark matter component. Investigating the behaviour and scaling relations of cored density profiles (which are the ground state of the wave function and often referred to as solitons) in more general scenarios is crucial to establish if an internal tension with FDM exists when considering the full phenomenology of the axiverse. At the higher mass end, the lack of detection of black hole super-radiance constrains ultralight masses between $10^{-19}\;\mathrm{eV}\lesssim m \lesssim 10^{-16}\;\mathrm{eV}$ and $10^{-13}\;\mathrm{eV}\lesssim m \lesssim 10^{-11}\;\mathrm{eV}$~\cite{Arvanitaki2011ExploringThe,Stott2018BlackHole}. The latter constraints still hold in the presence of multiple axion fields.

Axions are described by a scalar field $\phi$ with mass $m$ obeying the Klein-Gordon equation~\cite{Marsh2016AxionCosmology, Hui2017UltralightMatter}
\begin{align}
    \Box\phi-\frac{\partial V_\phi}{\partial \phi} =0,
\end{align}
where the box denotes the d'Alembertian operator and $V_\phi$ is the field potential. The general form of this potential is the periodic
\begin{align}
    V_\phi = m^2f_a^2\left[ 1-\cos\left(\frac{\phi}{f_a}\right)\right],
\end{align}
where $f_a$ is the field decay constant. In the limit of small displacements, $\phi\ll f_a$, the potential can be expressed as a quadratic: $V_\phi = m^2 \phi^2/2$. In the non-relativistic limit, the field has the ansatz~\cite{Widrow1993UsingMatter}
\begin{align}
    \phi = \frac{\hbar}{m\sqrt{2}}\left(\psi e^{-imt/\hbar}+\psi^* e^{imt/\hbar}\right),\label{eq:KW}
\end{align}
where $|\psi|^2$ is proportional to the FDM density in that limit.

%% METHODOLOGY %%

\section{Mixed Dark Matter Simulations}
\subsection{Wave Function Evolution}
Following the non-relativistic approximation of \cref{eq:KW}, the dynamics of FDM are governed by the Schrödinger equation
\begin{align}
    i\hbar \frac{\partial \psi}{\partial t}  = \left(-\frac{\hbar^2}{2ma^2} \nabla^2 + m\Phi_{\rm N} \right)\psi\,, \label{eq:Schrodinger}
\end{align}
where $m$ is the FDM particle mass and $\Phi_{\rm N}$ is the Newtonian gravitational potential. In a mixed dark matter cosmology, the evolution of the FDM component is coupled to CDM via the Poisson equation
\begin{align}
    \nabla^2 \Phi_{\rm N} = \frac{4\pi G}{a} \left(\rho_\mathrm{CDM} + \rho_\mathrm{FDM} - \bar{\rho}_\mathrm{tot}\right)\,, \label{eq:Poisson-combined}
\end{align}
where $\rho_\mathrm{FDM} = |\psi|^2$ and $\bar{\rho}_\mathrm{tot}$ is the mean of the total dark matter density.

To solve the above system of equations, we use \textsc{AxioNyx}~\cite{Schwabe2020SimulatingMixed} which is an extension of the cosmological simulation code \textsc{Nyx}~\cite{Almgren2013Nyx}.  
While the $N$-body solver of the \textsc{Nyx} code is used to evolve the CDM component, the pseudo-spectral solver described in Ref.~\cite{Schwabe2020SimulatingMixed} is used for the FDM component. 
This method has been shown to solve the Schr\"odinger-Poisson system accurately and to resolve the small-scale features of the wave function~\cite{Schwabe2020SimulatingMixed}. 
However, the latter is only possible if the simulation grid spacing is sufficiently small to resolve the de Broglie wavelength throughout the simulation. This means that the Schr\"odinger-Poisson solver is subject to the time-step criterion~\cite{Mocz2017GalaxyFormation,Edwards2018PyUltraLight}
\begin{align}
    \Delta t \leq 4\min\left[ \frac{m}{\hbar}\frac{(\Delta x)^2}{2\pi}, \frac{\pi \hbar}{4m|\Phi_{\rm N}|_\mathrm{max}}\right]\,, \label{eq:timestep} 
\end{align}
because the time step needs to capture the coherence time scale $t_c\sim (\Delta x)^2m/\hbar\sim \hbar/(mv^2)$ of the field. The constraints in both spatial and temporal resolution are the reason why FDM simulations cannot reach the spatial extent of pure CDM $N$-body simulations and most FDM simulations are stopped around redshift $z\lesssim 3$.

Multiple algorithms have been adapted or modified to solve the Schr\"odinger-Poisson system of equations. \textsc{Gamer}~\cite{Schive2014CosmicStructure} was notably one of the first followed by \textsc{AREPO}~\cite{Mocz2017GalaxyFormation,May2023TheHalo}, \textsc{ENZO}~\cite{Li2019NumericalAnd}, \textsc{NYX}~\cite{Schwabe2016SimulationsOf}, \textsc{GIZMO}~\cite{Hopkins2019AStable} and \textsc{RAMSES}~\cite{Mina2020ScalarAn}. A Python-based pseudo-spectral solver named \textsc{PyUltralight}~\cite{Edwards2018PyUltraLight} has also been recently developed. Some codes have been extended to simulate multiple axion fields~\cite{Guo2021TwoScalar,Glennon2023SimulationsOf,Gosenca2023MultifieldUltralight,Luu2023NestedSolitons}. There are also algorithms which take advantage of the fact that the Schr\"odinger-Poisson system admits a steady-state solution at the center of virialized structures. These express the wave function as a sum of eigen-functions which can be evolved forward in time at a reduced computational cost~\cite{Zagorac2023SolitonFormation}.

We exploit the efficiency of the pseudo-spectral solver to numerically evolve the FDM component and, given the size of the boxes and resolution used in this work, we stop the evolution of the wave function at redshift $z=4$ before all the modes become nonlinear. Since the pseudo-spectral solver relies on periodic boundary conditions, it cannot be used on higher levels of refinement where instead, a finite-difference solver is generally used. Finite-difference is used to approximate the Laplacian appearing in the Schr\"odinger-Poisson equations while, in the pseudo-spectral method, the Laplacian is computed without numerical approximations. This difference in the interplay of the two solvers can result in the lagging of the finite-difference solver with respect to the pseudo-spectral solver. For this reason, we postpone the implementation and testing of adaptive mesh refinement (AMR) in mixed dark matter simulations to future work.

%% ICs %%

\subsection{Initial Conditions}

\begin{figure}[h!]
    \centering
    \includegraphics[width=\linewidth]{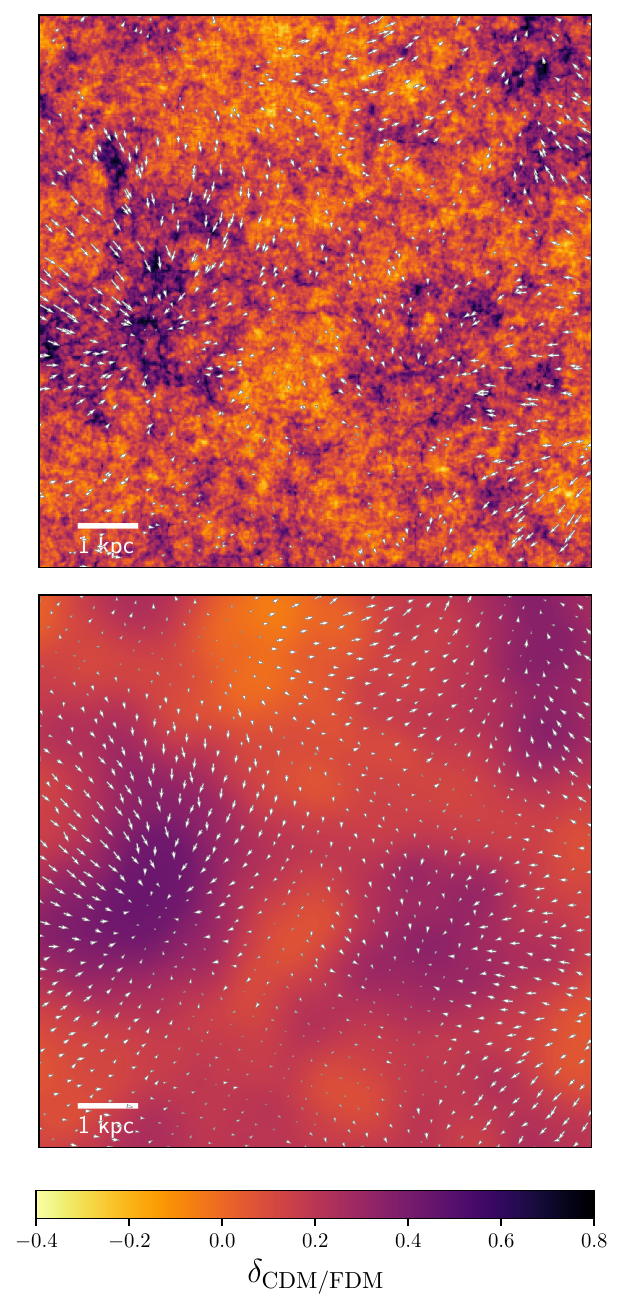}
    \caption{Initial density field for the CDM (top) and FDM (bottom) components of our simulation box 3 (see Table~\ref{tab:sim_list}) at $z=100$. The initial velocity fields are plotted and rescaled to the maximal velocity for each component. The ratio of the FDM to CDM velocity magnitudes goes to one on large scales.}
    \label{fig:ICs}
\end{figure}

%Initial conditions for systems with multiple components such as DM and baryons or in this case mixed DM 
Given the FDM wave effects, the CDM and FDM components will not have the same initial distributions of densities and velocities. 
One also has to use higher-order perturbation theory to avoid the formation of transient features in the simulations~\cite{Crocce2006TransientsFrom}. A second-order Lagrangian perturbation theory (LPT) scheme for mixed ultralight axions was developed in Ref.~\cite{Lague2021EvolvingUltralight}. The scale-dependent suppression in the FDM growth factor reproduces the effects of the wave diffusion at high redshifts since wave effects are captured at the linear level by the effective sound speed~\cite{Hwang2009AxionAs,Park2012AxionAs}
\begin{align}
    c^2_{s,\mathrm{eff}} = \frac{\hbar^2k^2}{4m^2a^2},
\end{align}
where $k$ is the comoving Fourier mode and $a$ is the cosmological scale factor. The modified LPT approach is, therefore, well suited for generating initial conditions for FDM cosmological simulations at high redshift (here we choose $z_\mathrm{ini}=100$ to be well into the linear regime). The main difference for the FDM component is that the velocity field at first order obeys
\begin{align}
    \mathbf{v}(\mathbf{k})=-\mathcal{H} \frac{d\ln D(k,a)}{d\ln a} \frac{\mathbf{k}}{k^2}\delta(\mathbf{k}), \label{eq:velocity_fourier}
\end{align}
where $\mathcal{H}=aH$, $\delta(k)$ is the Fourier transform of the overdensity field $\delta \equiv \rho/\bar{\rho}-1$, and $D$ is the linear growth factor of the density perturbations (for the CDM component, the growth factor is scale-independent). The density field used is to compute the velocities is the \textit{total} density, including both components. Therefore, the presence of FDM also affects the initial velocity of CDM particles by partially suppressing the initial power spectrum on small scales. In Ref.~\cite{Lague2021EvolvingUltralight}, it was shown that one can approximate
\begin{align}
    D(k, a) \approx L(k)D_\mathrm{CDM}(a),
\end{align}
where the prefactor $L$ is a monotonically decreasing function which asymptotically tends to zero on small sales. We use the public code \textsc{MUSIC}~\cite{Hahn2011Multi-Scale} along with the modified Boltzmann code \textsc{axionCAMB}~\cite{Hlozek2015AData}. We create a transfer function input file for each of the components and then generate initial positions and velocities for each, given the same total gravitational potential. The suppressed FDM transfer function ensures that the FDM particles are not given inconsistent velocities. The resulting initial density and velocity fields for both the CDM and FDM components of the simulations are shown for $f=0.1$ and $m=2.5\times 10^{-22}$ eV in Fig~\ref{fig:ICs}. The difference in the number and placements of the velocity field arrows is due to the fact that the velocity field of the FDM is evaluated on a grid rather than with particles. To initialize the wave function, we use the Madelung change of variables $\psi=Re^{i\theta}$ where the magnitude and phase can be obtained with
\begin{align}
    R(\mathbf{x}) &\equiv \sqrt{\frac{\rho_\mathrm{FDM}(\mathbf{x})}{m}},\\
    \nabla \theta(\mathbf{x}) &\equiv \frac{m}{\hbar} \mathbf{v}_\mathrm{FDM}(\mathbf{x}). \label{eq:fdm_vel_ic}
\end{align}
To obtain the phase of the wave function, Eq.~(\ref{eq:fdm_vel_ic}) can be solved in Fourier space given the Fourier transform of the velocity field which was calculated with Eq.~(\ref{eq:velocity_fourier}). In Fig.~\ref{fig:ICs}, we can observe the difference in clustering between the two species at high redshift. Many simulations have used the modified initial conditions of FDM combined with CDM-like evolution to approximate the behaviour of FDM at a lower computational cost (by not solving the Schr\"odinger equation). This has been known as the warm dark matter (WDM) or classical FDM approach~\cite{Mocz2019FirstStar, Dome2023OnThe}. We will test the validity of this approximation for low FDM concentrations in Sec.~\ref{sec:results}.

%% RESULTS %%

\section{Results \label{sec:results}}
We simulate the evolution of the CDM and FDM components from the initial redshift $z_\mathrm{ini}=100$. We generate a set of 12 simulation boxes listed in Table~\ref{tab:sim_list}. First, we fix the mass at $m=2.5\times 10^{-22}$ eV and select five FDM fractions: $f\in \{0, \; 0.01,\; 0.1, \;0.5, \;1\}$. We then consider four other masses $m\in 2.5\times\{10^{-25},\; 10^{-24}, \;10^{-23}, \;10^{-21}\}$ eV, but keep the fraction at $f=0.1$ for these runs. We also create two simulations where the initial conditions account for 1\% and 10\% FDM, but the system is evolved using only CDM particles and a simulation with twice the resolution using a grid of $1024^3$ points for a convergence analysis (and Appendix~\ref{app:resolution}). Snapshots of the total matter density in boxes 1, 4, and 5 at redshift $z=4$ are displayed in Fig.~\ref{fig:density_slice}. The visualizations are generated using the analysis code \textsc{yt}~\cite{Turk2011YT}.
\renewcommand{\arraystretch}{1.5}%
\begin{center}
    \begin{tabular}{ |c|c|c|c|c| } 
        \hline
        Box & FDM & FDM & Box & $N$ Grid/ \\%& $z_\mathrm{final}$\\ 
        Number & Mass (eV) & Fraction & Length & Particles \\
        \hline
        1 & $2.5\times 10^{-22}$ & $0.00$ & 1.0 Mpc$/h$ & $512^3$ \\%& 4\\ 
        2 & $2.5\times 10^{-22}$ & $0.01$ & 1.0 Mpc$/h$ & $512^3$ \\%& 4\\ 
        3 & $2.5\times 10^{-22}$ & $0.10$ & 1.0 Mpc$/h$ & $512^3$ \\%& 4\\ 
        4 & $2.5\times 10^{-22}$ & $0.50$ & 1.0 Mpc$/h$ & $512^3$ \\%& 4\\ 
        5 & $2.5\times 10^{-22}$ & $1.00$ & 1.0 Mpc$/h$ & $512^3$ \\%& 4\\ 
        6 & $2.5\times 10^{-21}$ & $0.10$ & 0.3 Mpc$/h$ & $512^3$ \\%& 4\\ 
        7 & $2.5\times 10^{-23}$ & $0.10$ & 1.0 Mpc$/h$ & $512^3$ \\%& 4\\ 
        8 & $2.5\times 10^{-24}$ & $0.10$ & 10.0 Mpc$/h$ & $512^3$ \\%& 4\\
        9 & $2.5\times 10^{-25}$ & $0.10$ & 30.0 Mpc$/h$ & $512^3$ \\%& 4\\
        10 & $2.5\times 10^{-22}$ & $0.10$ & 1.0 Mpc$/h$ & $1024^3$ \\%& 4\\
        11 & FDM ICs & 0.01 & 1.0 Mpc$/h$ & $512^3$ \\
        12 & FDM ICs & 0.10 & 1.0 Mpc$/h$ & $512^3$ \\
        \hline
    \end{tabular}
    \captionof{table}{List of completed simulations with FDM masses and fractions. The box size is given in comoving coordinates. The number of particles for the CDM component is the same as the number of grid points for the FDM in all cases. FDM ICs under particle mass indicates that the particles were evolved as CDM with modified initial conditions.
    \label{tab:sim_list}}
\end{center}

\subsection{Non-Linear Matter Power Spectrum}
\begin{figure*}
    \centering
    \includegraphics[width=\linewidth]{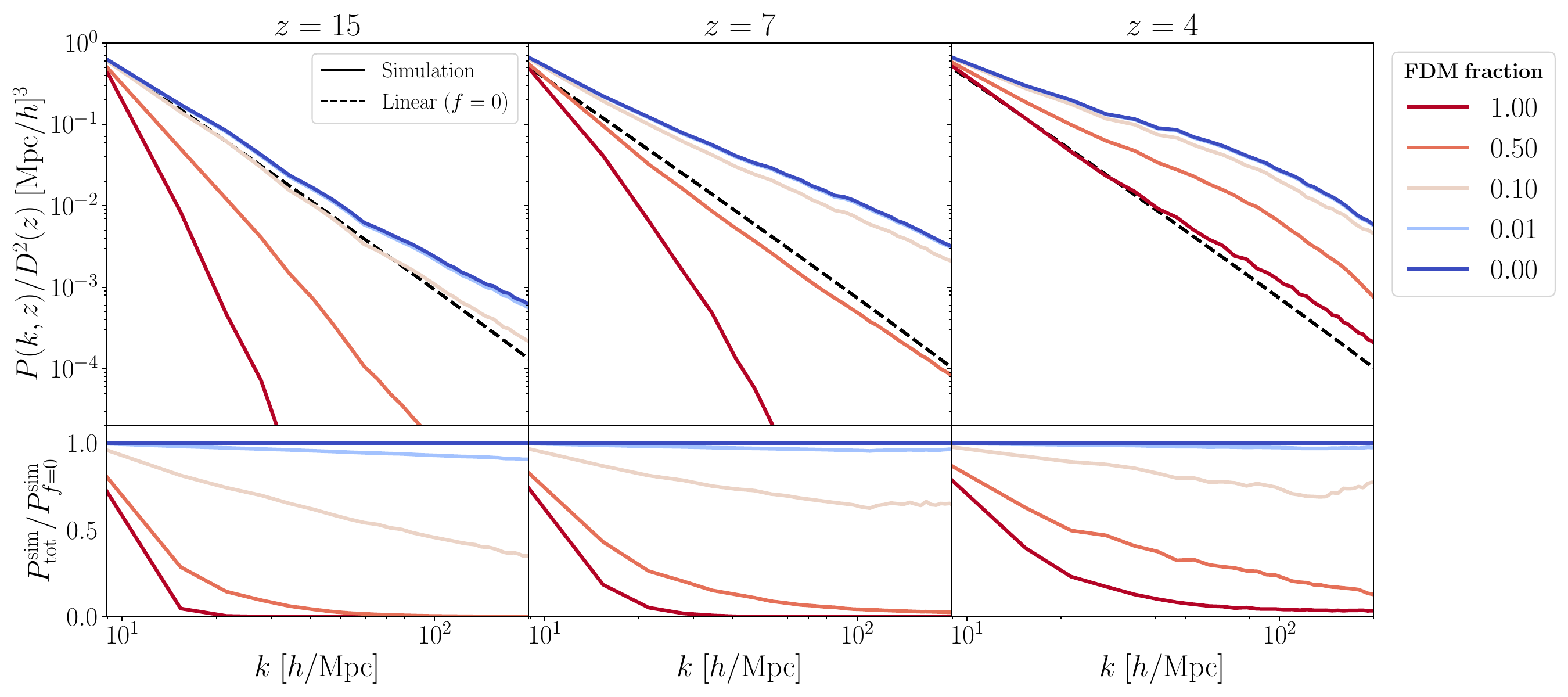}
    \caption{Power spectra in the simulation boxes as a function of redshift and FDM fraction. (Top) Combined (CDM+FDM) power spectra normalized by the time-dependent growth factor. The dashed line represents the prediction from linear theory for pure CDM. (Bottom) Ratio of the combined power spectra with respect to the non-linear pure CDM power spectrum at the same redshift.
    \label{fig:nlmps}}
\end{figure*}
\begin{figure*}
    \centering
    \includegraphics[width=\linewidth]{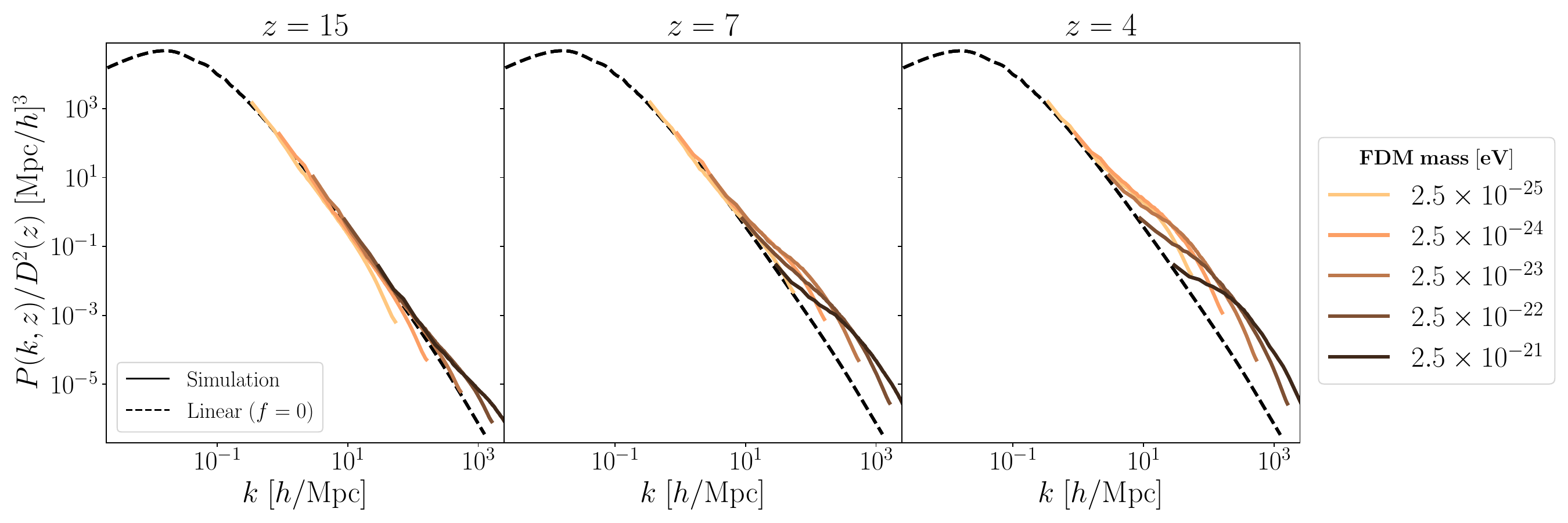}
    \caption{Combined (CDM+FDM) power spectra normalized by the time-dependent growth factor in the simulation boxes as a function of redshift and FDM mass for a fraction of $f=0.1$. We note the formation of non-linear structures at redshift $z=7$ for all masses which is not found in simulations where $f=1$.
    \label{fig:nlmps_mass}}
\end{figure*}

We extract the total power spectra from our simulations using the public code \textsc{nbodykit}~\cite{Hand2017nbodykit}.
The spectra and relative difference of the FDM and CDM power spectra for a series of FDM fractions is shown as a function of redshift in Figs~\ref{fig:nlmps}-\ref{fig:nlmps_mass}. We have factored out the homogeneous (large-scale) growth factor $D(z) \propto 1/(1+z)$ in the matter-dominated Universe to highlight the growth of structure on small scales. For high FDM fractions (redder curves), the FDM and CDM spectra are relatively close together at high redshift. In this case, the CDM follows the distribution of the FDM and will be similarly suppressed on small scales. However, the converse does not hold. If the CDM is the main component of the dark matter, the FDM still does not cluster on scales below its Jeans scale (where the FDM sound speed equates to the gravitational potential)~\cite{Khlopov1985GravitationalInstability,Marsh2016AxionCosmology}, which in the pure FDM case is given by 
\begin{align}
    k_{\rm J} = 66.5 a^{1/4}\left(\frac{m}{10^{-22}\;\mathrm{eV}}\right)^{1/2} \left(\frac{\Omega_\mathrm{FDM}h^2}{0.12}\right)^{1/4}\;\mathrm{Mpc}^{-1}.
\end{align}
This is visible in the pale blue and grey curves in the $z=15$ panel of Fig.~\ref{fig:nlmps}.

In the bottom row of Fig.~\ref{fig:nlmps}, we also plot the ratio of the total matter power spectrum to the non-linear matter power spectrum of the pure CDM case ($f=0$). As observed in Ref.~\cite{Lague2022ConstrainingUltralight}, even a FDM fraction of a few percents leads to a strong suppression of clustering. This explains why the 10\% FDM fraction (pink line) is damped by over 50\% at redshift $z=15$. We note however, that the FDM rapidly falls into the CDM potential well, and the total non-linear power spectrum approaches that of the pure CDM case as structure forms at lower redshifts. Comparing the results shown in Fig.~\ref{fig:nlmps_mass} with the simulations of Ref.~\cite{May2021StructureFormation}, we can see that non-linearities form sooner in the $f=0.1$ case than in the $f=1$ case as a power spectrum excess (compared to the linear prediction) is visible at redshift $z\sim7$. This is mostly attributable to the lower level of suppression in the linear spectrum for $f<1$, but also to the scale-dependent structure growth of FDM which delays the in-fall of FDM in halos~\cite{Lague2021EvolvingUltralight}.

\begin{figure}
    \centering
    \includegraphics[width=\linewidth]{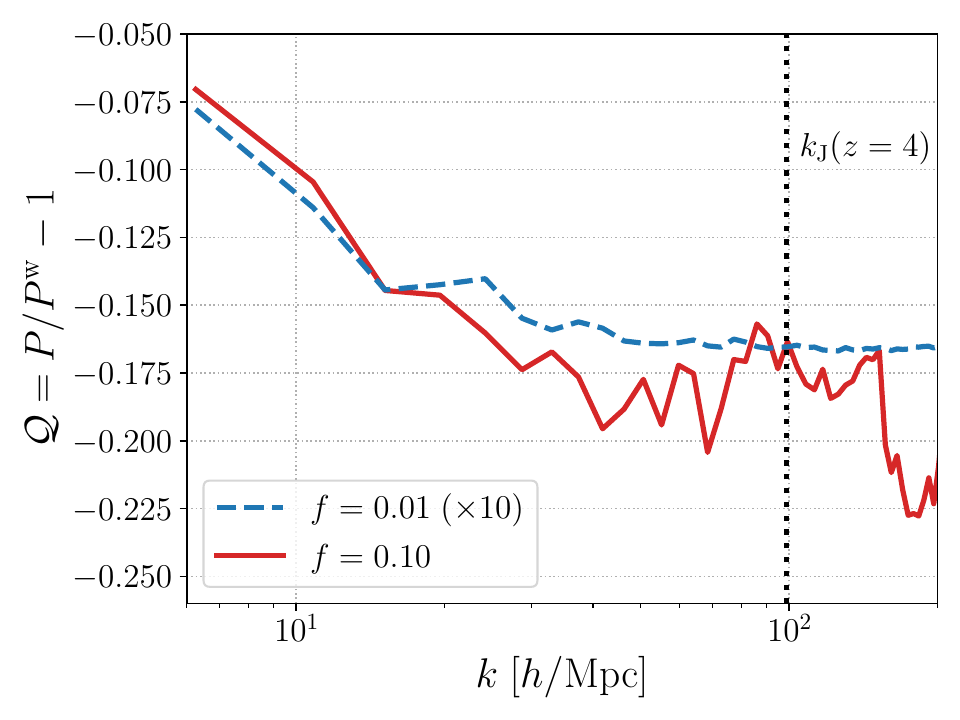}
    \caption{Impact of wave effects on the power spectrum at redshift $z=4$. We increase the amplitude of $\mathcal{Q}$ by an order of magnitude for $f=0.01$ to compare the shape of the curves across FDM fractions. The late-time FDM wave effects not captured by the modified initial conditions are more apparent when $f=0.1$. The dotted black line denotes the Jeans scale for a mass of $2.5\times 10^{-22}$ eV.}
    \label{fig:WDM-FDM}
\end{figure}
To measure the impact of the wave effects and differentiate them from the structure suppression caused by the modified initial conditions, we run two additional simulations using the WDM approximation as used in Refs~\cite{Mocz2019FirstStar,Dome2023OnThe}. The WDM approach consists in modifying the initial conditions to account for the presence of FDM but to evolve the system for $z<z_\mathrm{ini}$ using only the CDM dynamics. For this, we use the initial conditions of boxes 2 and 3 with $f=0.01-0.10$, we label the power spectra of such simulations $P^\mathrm{w}_f$. Taking the ratio of this power spectrum to the power spectrum of the simulation boxes which account for both the modified initial conditions \textit{and} the FDM dynamics, we have
\begin{align}
    \mathcal{Q}(k) \equiv \frac{P_{f}(k)}{P^\mathrm{w}_{f}(k)}-1.
\end{align}
The role of the function $\mathcal{Q}(k)$ is to isolate the impact of the late-time effect of the wave behaviour of FDM. If this function is exactly zero, this tells us the wave effects play no role in clustering and that the all the suppression observed in Figs~\ref{fig:nlmps}-\ref{fig:nlmps_mass} is due solely to the initial conditions. We observe increased suppression of the non-linear matter power spectrum when accounting for the wave effects, as shown in Fig.~\ref{fig:WDM-FDM}. We note that the shape of the function $\mathcal{Q}$ remains the same as we increase the FDM fraction, and only its amplitude changes. Our results agree with the finding of Ref.~\cite{Nori2018Ax-Gadget} with the added property that difference between the WDM and FDM treatments becoming negligible at low fractions. In other words, we find $\mathcal{Q}(k)\to 0$ as $f\to 0$ roughly linearly.

\subsection{Density Profiles}
A well-known prediction of FDM models is the formation of solitonic cores at the center of halos~\cite{Schive2014CosmicStructure}. We can obtain the shape of this core by noting that the Schr\"odinger-Poisson system has an equilibrium solution with a complex phase. The wave function takes the form $\psi(\mathbf{x}, t) = e^{-i\gamma t/\hbar} \phi(\mathbf{x})$, where $\gamma$ is a constant. Assuming spherical symmetry (with $r=|\mathbf{x}|$), the system of equations~(\ref{eq:Schrodinger})-(\ref{eq:Poisson-combined}) becomes~\cite{Guzman2004EvolutionOf,Veltmaat2020Baryon-Driven}
\begin{align}
    \frac{\partial^2 (r\phi)}{\partial r^2} &= 2r\left(\frac{m^2}{\hbar^2}V-\frac{m}{\hbar^2}\gamma\right) \phi, \\
     \frac{\partial^2 (rV)}{\partial r^2} &=4\pi G r\left(\phi^2+\rho_\mathrm{CDM}\right).\label{eq:shooting}
\end{align}

We fit the density profile of the FDM by solving for the equilibrium solution given the shape of the CDM density profile and the ratio of their central densities. We develop a more stable alternative to the shooting method used in previous studies to find the eigenvalue $\gamma$~\cite{Guzman2004EvolutionOf} which we describe in Appendix~\ref{app:pade}. We confirm that the equilibrium state of the scalar function $\phi$ does satisfy the Schr\"odinger-Poisson system but with a different eigenvalue than the one found for the zero-node solution in the pure FDM case.
\begin{figure}[t]
    \centering
    \includegraphics[width=\linewidth]{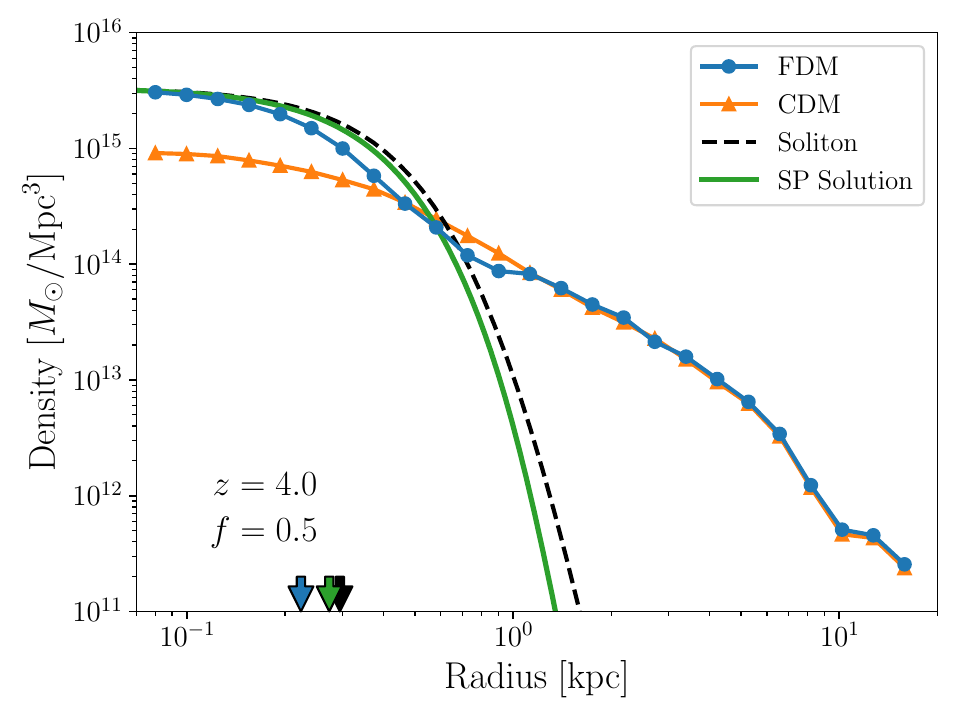}
    \caption{CDM and FDM density profiles in simulation box 4 with 50\% FDM (see Table~\ref{tab:sim_list}). The dark dashed line represents a traditional soliton profile fit to the FDM density and the green line denotes the fit found by solving Eq.~(\ref{eq:shooting}) in the presence of CDM. The arrows denote the half-density radius of the curves of the corresponding color.}
    \label{fig:shooting-method}
\end{figure}

\begin{figure}[t]
    \centering
    \includegraphics[width=\linewidth]{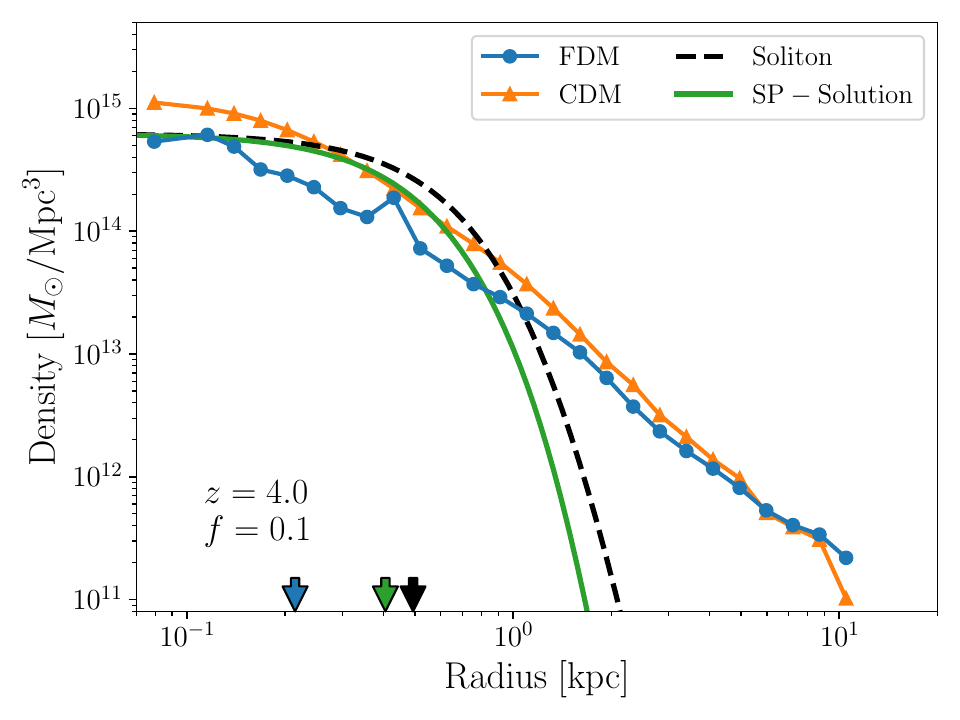}
    \caption{CDM and FDM density profiles in simulation box 3 with 10\% FDM (see Table~\ref{tab:sim_list}). The dark dashed line represents a traditional soliton profile fit to the FDM density and the green line denotes the fit found by solving Eq.~(\ref{eq:shooting}) in the presence of CDM. The arrows denote the half-density radius of the curves of the corresponding color.}
    \label{fig:density_010}
\end{figure}

As with a baryon-FDM mixture, we observe that a secondary contributor to the potential of the FDM causes the core radius to shrink~\cite{Veltmaat2020Baryon-Driven}. In Fig.~\ref{fig:shooting-method} and Fig.~\ref{fig:density_010}, we display the CDM and FDM density alongside the canonical soliton fit of Ref.~\cite{Schive2014CosmicStructure}. The fit can be obtained with two free parameters giving
\begin{align}
    \rho_\mathrm{sol}(r) = \frac{\rho_0}{\left[1+0.091(r/r_c)^2\right]^8},
\end{align}
where $\rho_0$ is the central density and where $r_c$ is the radius at which the density reaches half of its central value. The FDM mass and the two parameters of the fit obey the scaling relation
\begin{align}
    r_c = 1.0&\left(\frac{\rho_0}{3.1\times 10^{15}\;M_\odot/\mathrm{Mpc}^3}\right)^{1/4} \nonumber\\&\times \left(\frac{m}{2.5\times10^{-22}\;\mathrm{eV}}\right)^{1/2} \;\mathrm{kpc}.
\end{align}
We find in the mixed CDM-FDM case that the relation for the core radius no longer holds. In the case where $f=0.5$, the FDM dominates the inner region of the halo and the soliton fit still provides a good description of the inner FDM density with a slightly different slope at high $r$. However, when CDM composes 90\% of the dark matter, we do not detect a proper soliton core. We test to see if the core radius falls below the resolution of the simulation by solving the time-independent ground state of the Schr\"odinger-Poisson system in the presence of CDM. The CDM creates a steeper gravitational potential gradient on the FDM wave function causing it to be radially compressed. In our simulations, we find that the FDM density profile's inner slope is too steep to be well modeled by a proper soliton core for the range of radii we can resolve. Our results agree with previous studies on mixed dark matter using spherical collapse, which found no core formation for fractions below $f\lesssim 0.3$~\cite{Schwabe2020SimulatingMixed}.

On the outskirts of the halos, we find that the total density profile approaches the Navarro-Frenk-White (NFW) fit~\cite{Navarro1996TheStructure}:
\begin{align}
    \rho(r) = \frac{\rho_s}{(r/r_s)(1+r/r_s)^2},\label{eq:nfw-fit}
\end{align}
where $r_s$ is the NFW scale radius and $\rho_s$ is its characteristic density. The NFW profile has a diverging mass as $r\to \infty$, which is unphysical. It is customary to denote the limit of the halo with its virial radius, which is defined as a function of the virial overdensity $\Delta_\mathrm{vir}$ given by~\cite{Bryan1998StatisticalProperties}
\begin{align}
    \Delta_{\rm vir} = 18\pi^2 +82 \left[\Omega_\mathrm{m}(z)-1\right] -39 \left[\Omega_\mathrm{m}(z)-1\right]^2\label{eq:delta_vir}
\end{align}
and which we can relate to the halo mass through
\begin{align}
    M(r_{\rm vir})=\frac{4\pi}{3} r_{\rm vir}^3 \Delta_{\rm vir} \bar{\rho}_{\rm tot}.
\end{align}
We will use this definition in the following section.

\begin{figure*}
    \centering
    \includegraphics[width=0.85\linewidth]{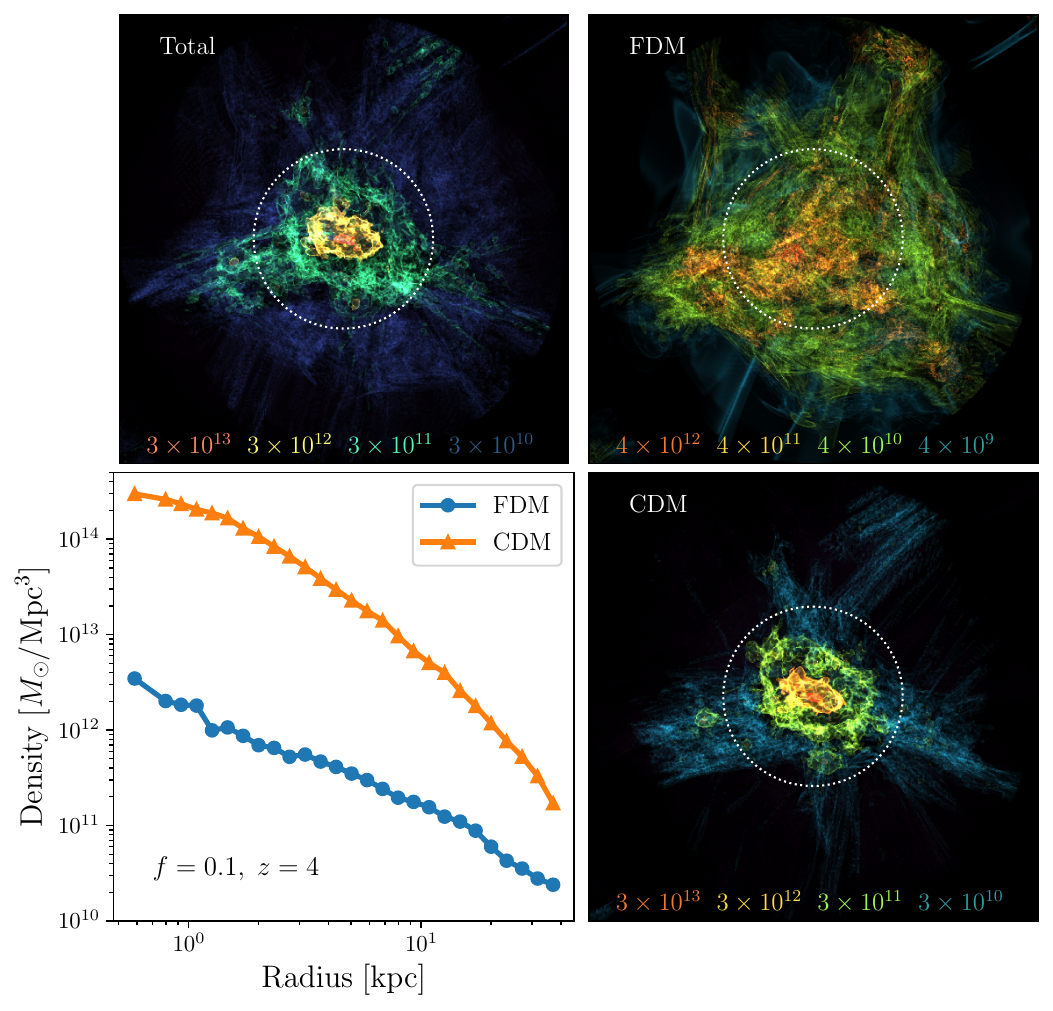}
    \caption{Tri-dimensional density isocontours around a CDM-dominated sub-halo with an FDM mass of $m=2.5\times 10^{-22}$ eV. Despite being in a simulation with $f=0.1$, $f\ll 0.1$ within the halo, indicating it is close to the cut-off where no FDM is clustered. The three isocontour panels show the separate components and the total density. The lower left panel shows the measured density of CDM and FDM as a function of the distance from the sub-halo center up to 40 kpc which corresponds to the white dashed circle in the other three panels. The colored text indicates the density of the isocontours in units of $M_\odot/\mathrm{Mpc}^3$.}
    \label{fig:cdm-dominated-halo}
\end{figure*}

A key assumption in the halo model of Ref.~\cite{Vogt2023ImprovedMixed} is that CDM could form bound structures which contain little to no FDM. This can happen when a halo has a radius smaller than the halo Jeans radius ($r_\mathrm{hJ}$) which is defined as the radius where
\begin{align}
    \rho(r_\mathrm{hJ}) \approx \frac{\bar{\rho}_{\rm tot} \Delta_{\rm vir} c^2 r_{\rm vir}}{3f(c)r_\mathrm{hJ}}, \label{eq:rhj}
\end{align}
where $c$ is the concentration of the halo which is defined as the ratio of the virial radius to the scale radius (from the NFW fit) $c\equiv r_{\rm vir}/r_s$ and $f(x) = \ln(x+1) - \frac{x}{x+1}$. However, Eq.~\ref{eq:rhj} is only valid in the limit where $r_{\rm vir}\geq r_\mathrm{hJ}$. For a FDM particle mass of $2.5\times 10^{-22}$ eV at $f=0.1$, this translates to a halo mass of about $10^{6}\;M_\odot/h$. Halos for which this does not hold are taken to be devoid of FDM in the halo model of Ref.~\cite{Vogt2023ImprovedMixed}. 

In Fig.~\ref{fig:cdm-dominated-halo}, we show a halo of simulation box 3 and its surrounding region. By plotting the density isocontours, we notice that the FDM component is very diffuse while the CDM has a steep density gradient. This is shown quantitatively in the lower left panel of Fig.~\ref{fig:cdm-dominated-halo} where the CDM is more than 100 times denser in the center of the region than the FDM. When calculating the Jeans radius for this halo, we find no value of $r_\mathrm{hJ}$ satisfying Eq.~(\ref{eq:rhj}) and conclude that $r_{\rm vir}< r_\mathrm{hJ}$. The findings shown in Fig.~\ref{fig:cdm-dominated-halo} supports the approximations of the halo model as the halo is composed at $\approx 97.5$\% of CDM.

\begin{figure}
    \centering
    \includegraphics[width=\linewidth]{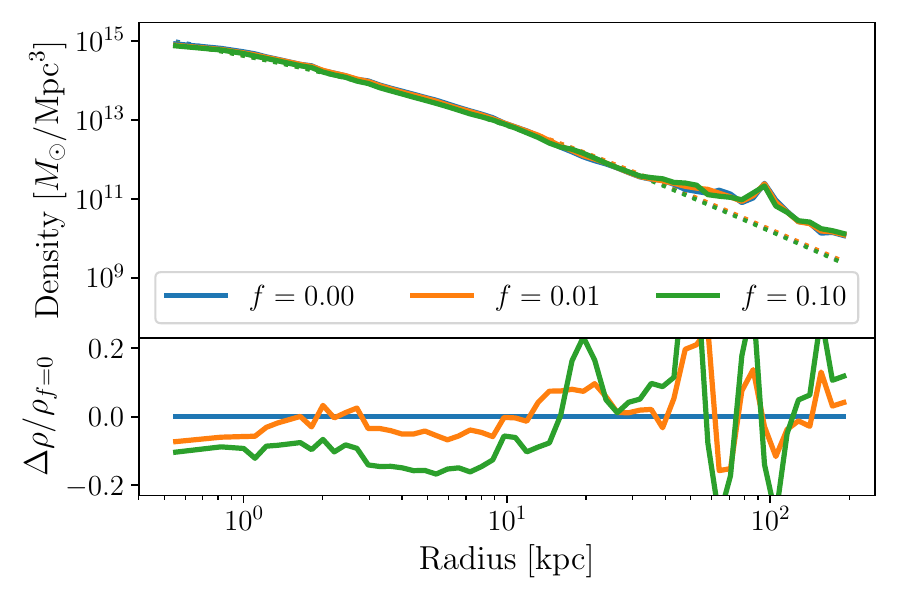}
    \caption{(\textit{Top}) Radial density profile of the most massive halos in low FDM fraction simulation boxes for a FDM mass of $m=2.5\times 10^{-22}$ eV and a redshift $z=4$. The dotted lines represent the NFW profile fits to the total density. (\textit{Bottom}) Relative difference in density with respect to the CDM-only ($f=0$) case.}
    \label{fig:low-frac-stacked}
\end{figure}

In the limit of high halo mass with low FDM fraction, we find that the total (combined) density is largely unaffected by the presence of FDM. This is shown for fractions $f\leq 0.1$ in Fig.~\ref{fig:low-frac-stacked}. We fit the radial density of the most massive halo in each of the boxes 1-3 using a NFW profile and found the concentration parameter to be unchanged by small amounts of FDM even if the halo central density is slightly reduced. Moreover, the FDM fraction calculated within the three halos of Fig.~\ref{fig:low-frac-stacked} are $(0.000, 0.107, 0.011)$ for cosmological FDM fractions $f=(0.00, 0.10, 0.01)$, respectively. Thus, we find that massive halos with a radius larger than the Jeans radius accumulate a concentration of FDM equal to the cosmological FDM fraction. This matches the prediction of the biased tracer model of Ref.~\cite{Vogt2023ImprovedMixed}.

\subsection{Halo Mass Function}
In this section, we measure the halo abundance and compare our results with the theoretical predictions of a halo model for mixed dark matter. In a mixture of CDM and FDM, the CDM is free to coalesce in small halos while the FDM substructure is still washed out by wave effects. Therefore, unlike cosmologies with a FDM fraction of $f=1$, halos with a radius below the FDM Jeans scale will still form in mixed dark matter. To investigate this quantitatively, we measure the halo mass function (HMF) in simulation boxes 1 and 3 (see Table~\ref{tab:sim_list}).

We use the CDM particles as a tracer of the halos and use the friends-of-friends (FOF) algorithm~\cite{Davis1985TheEvolution} to identify the halo centers. To correct for the presence of FDM, we create a spherical profile around the halo center and vary its radius until the total mean density of the halo reaches the virial overdensity given in Eq.~(\ref{eq:delta_vir}). We only use this approach in simulations where CDM composes at least 90\% of the dark matter and where FDM acts as a tracer of the CDM. The CDM particles have a smaller separation between them in high density regions, and not all halos that are resolved with particles have a high enough number of grid cells to apply the FDM mass correction. About 80 halos had a radius sufficiently large to accurately account for the presence of FDM. The final halo masses used are found by summing over the mass of the particles in the halos and applying the FDM halo mass correction obtained from the spherical profile. The other halos identified by the FOF algorithm had a radius smaller or comparable to the FDM de Broglie wavelength. One such halo which had a sufficiently large radius to encompass many FDM grid cells is shown in Fig.~\ref{fig:cdm-dominated-halo}. Most of the halos, however, did not have a sufficient number of grid points to resolve the FDM density profile and the FDM mass correction assumed a flat density profile equal to the mean FDM density.

\begin{figure}[b]
    \centering
    \includegraphics[width=\linewidth]{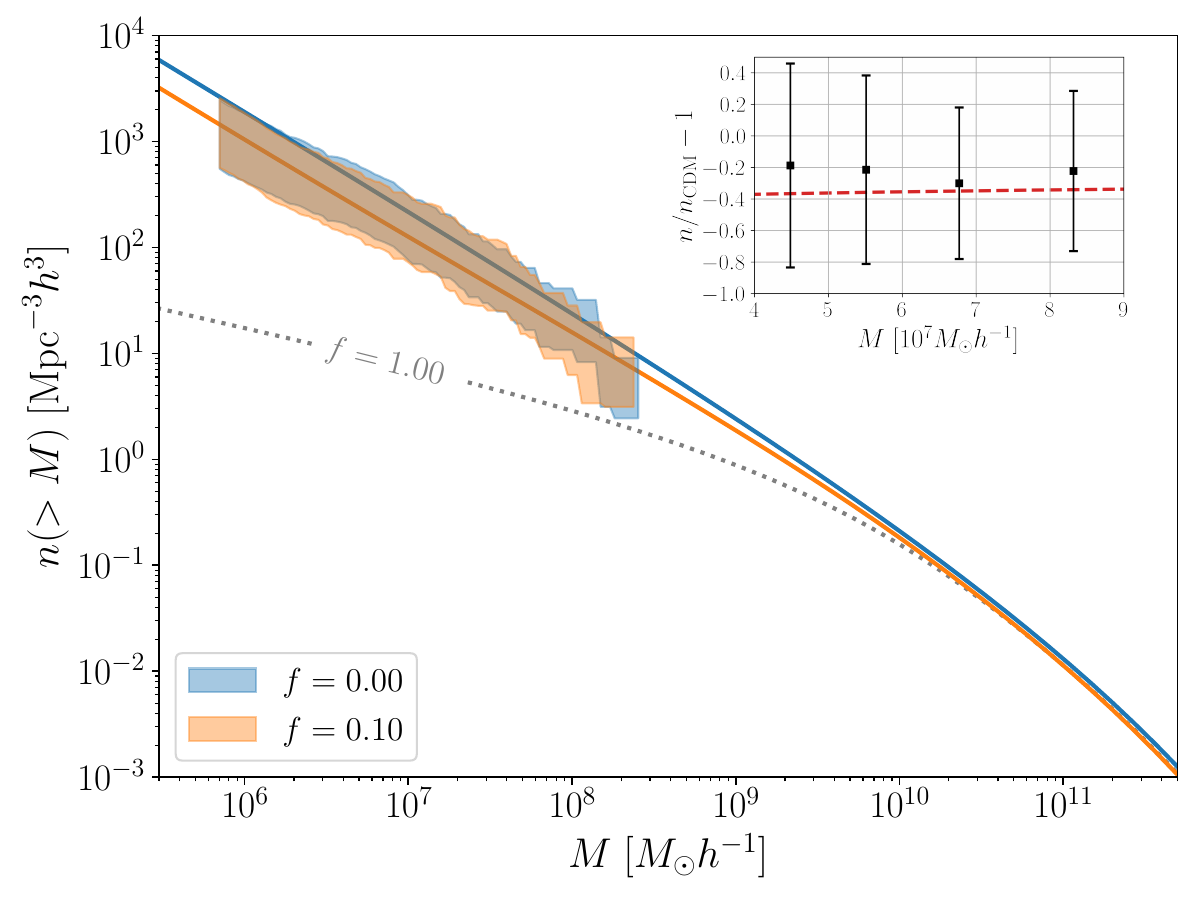}
    \caption{Halo mass function (number density of halos above a certain mass threshold) for two FDM fractions at redshift $z=4$. The colored regions are measurements from simulations with the 68\% confidence intervals and the solid lines correspond to the predictions from \textsc{axionHMCode}. (\textit{Inset}) Relative difference in the number density of halos in the mass range $10^{7}-10^{8}\;M_\odot/h$ compared to the model prediction (red dashed).}
    \label{fig:hmf-comparison}
\end{figure}

We compare our results with the halo model predictions of \textsc{axionHMCode}~\cite{Vogt2023ImprovedMixed}, which is a an adaptation of \textsc{HMCode}~\cite{Mead2021HMCODE} accounting for ultralight particles composing part of the dark matter. It is based on the halo model first implemented in Ref.~\cite{Dentler2022FuzzyDark} to compute weak lensing shear statistics with FDM. We show our results along with the model prediction in Fig.~\ref{fig:hmf-comparison}. We split the simulation volume in four boxes and compute the error bars using a jackknife method. To improve the accuracy of the comparison, we also rescale the HMF prediction. This rescaling is necessary due to the fact that the HMF model assumes a different mass definition than the output of the FOF halo finder. We use the python package \textsc{hmf}~\cite{Murray2013HMFcalc, Murray2021THEHALOMOD} to compute the ratio of HMFs using the FOF and virial mass definitions (for a more in-depth discussion of halo mass definitions see~\cite{More2011TheOverdensity}). The rescaling is roughly mass-independent and has a value of $\sim 0.83$. This difference is negligible compared to the error margin on the HMF measurement and does not affect our results. %Given the small size of our box, we only detect about 100 halos, thus making the halo mass function measurement noisy. Nevertheless,
We find a reasonable agreement between the model and the simulations. We expect the halo mass functions for different FDM fractions to converge at halo masses larger than what can be captured by our box size as shown by the model prediction. Given our box size, the results for the HMF are somewhat inconclusive, although in the largest mass bins, the measured HMF shows a similar decrement to the theoretical HMF (see the inset of Fig.~\ref{fig:hmf-comparison}).

\subsection{Halo Shapes}
Halo triaxiality has been suggested as a way to detect dark matter properties beyond CDM, such as dark matter self-interactions~\cite{Dave2001HaloProperties,Peter2013CosmologicalSimulations,Gonzalez2023ClusterHalo}. In this section, we investigate the impact of FDM on the shape of dark matter halos. Given the presence of a spherical solitonic core, we expect that the FDM will lead to more spherical halo centers than pure CDM. Further away from the halo center, a suppression of the initial power spectrum can also give more spherical halos in the absence of wave effects as shown in Ref.~\cite{Dome2023OnThe}. To test this prediction, we compare the shape of halos for FDM fractions $f=\{0.00, \;0.01,\; 0.10,\; 0.50\}$ and use the central halo in simulation boxes 1, 2, 3, and 4. 

Halo shapes are parameterized by looking at the length of the halo's principal axes which are labelled in increasing size $a\geq b \geq c$. To compute the direction and relative length of the principal axes, we diagonalize the reduced inertia tensor $\mathcal{I}$ of the halo. The inertia tensor can be computed using different methods~\cite{Zemp2011OnDetermining}, but we follow the approach of Refs~\cite{Dubinski1991TheStructure, Dave2001HaloProperties, Allgood2006TheShape} giving
\begin{align}
    \mathcal{I}_{ij} = \sum_{n\; \rm in\;halo} \left[1+\delta_{\mathrm{tot}}(\mathbf{x}_n)\right]\frac{ x_{n,i} x_{n,j}}{r_n^2},
\end{align}
where $i,\;j=1,2,3$ are the indexes of the axes, $x_{n}$ is the distance from the center of the halo to the $n^{\rm th}$ grid point in the ellipsoid axis frame, $r_n=\sqrt{x_{n,1}^2+x_{n,2}^2/q^2+x_{n,3}^2/s^2}$ is the ellipsoidal radius. The ellipsoidal radius is obtained from the axes ratios $q=b/a$ and $s=c/a$. Normally, the inertia tensor is computed as a sum over particles of the same mass. In the case at hand, we take a weighted sum over grid points located within the central halo. We interpolate the particle mass density of the CDM using a cloud-in-cell scheme on the same grid as the FDM and we weigh the grid points by their total matter overdensity $\delta_{\rm tot} = f\delta_{\rm FDM}+(1-f)\delta_{\rm CDM}$. We then compute this tensor iteratively following the procedure of Ref.~\cite{Allgood2006TheShape}. We begin by calculating the principal axes for a halo centered at the point of highest density. The radius of the sphere is taken to be the virial radius. Then, we create an ellipsoid of the same volume as the initial sphere, with the principal axes aligned with the eigenvectors of the inertia tensor and with the proper axes ratios. The lengths of the axes simply correspond to the square root of the eigenvalues of the tensor. This ellipsoid defines a new halo for which we repeat the same procedure until we have reached ten iterations (after which the inferred shape of the halo converges). For a numerical implementation of this procedure, we make use of the public repositories \textsc{inertia-tensors}~\cite{Duncan2019InertiaTensors} and \textsc{rotations}~\cite{Duncan2020Rotations}.

To quantify the ellipticity of halos, we calculate the triaxiality parameter~\cite{Franx1991TheOrdered,Allgood2006TheShape, Schneider2012TheShapes}
\begin{align}
    T\equiv \frac{a^2-b^2}{a^2-c^2}.
\end{align}
Using this parameter, we can categorize halos as oblate ($T\leq 0.33$), prolate ($T\geq 0.66$), or triaxial ($0.33 \leq T \leq 0.66$). 
\begin{figure}
    \centering
    \includegraphics[width=\linewidth]{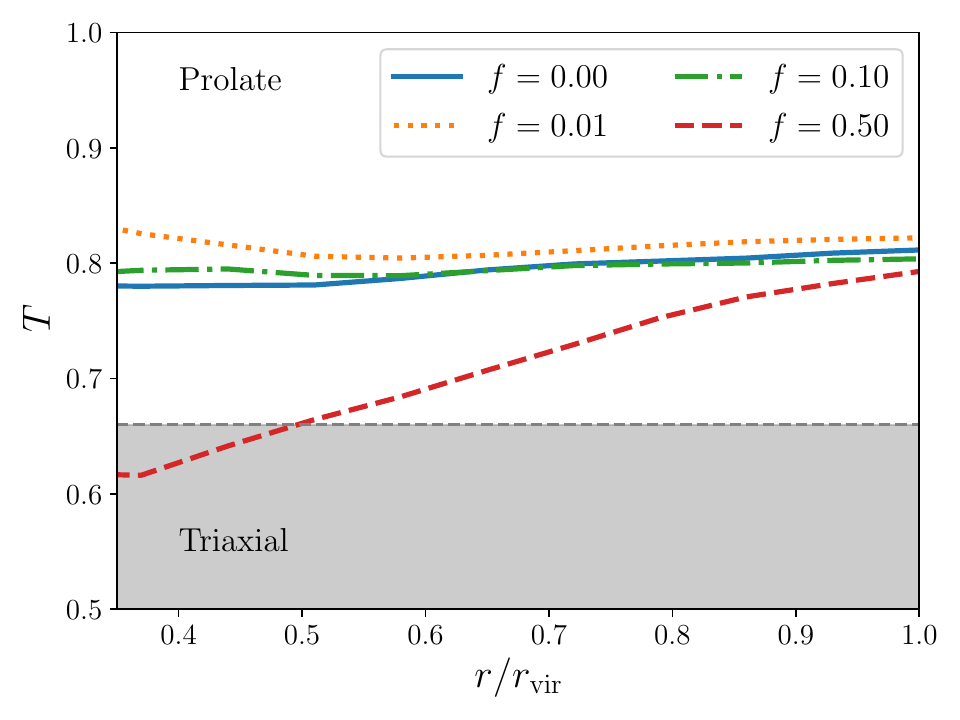}
    \caption{Triaxial parameter of the central halo for various FDM fractions. We note a more spherical inner halo region with high fraction due to the presence of the spherical soliton core. The $f=1$ halo was too diffuse and the ellipsoid algorithm did not converge.
    }
    \label{fig:t_param}
\end{figure}
In Fig.~\ref{fig:t_param}, we show the impact of having a high fraction at various radii. We note that the effects on halo shape only manifest at around $f=0.5$, where the central region within $0.5\;r_{\rm vir}$ is more spherical. Contrary to past studies on halo triaxiality, we are not comparing halo populations but identical halos in simulations with different dark matter contents. We also only study halos at high redshift $z=4$. Since the formation of a soliton core is universal in halos containing ultralight particles, we expect our observations to hold for halo populations at lower redshifts. We note that our algorithm did not converge for the FDM-only $f=1$ case as the structure was too diffuse and contained a significant number of interference fringes.

%% DISCUSSION/CONCLUSION %%

\section{Discussion}
In this study, we have explored the non-linear behaviour of dark matter composed of a mixture of cold and ultralight (fuzzy) dark matter. We adapted a simulation algorithm for dark matter models with gravitationally coupled CDM and FDM components evolved simultaneously.

We first examined the non-linear power spectrum of each component up to redshift $z=4$.
While the resulting spectra proved too noisy for a comparison with halo model predictions, we confirmed that the presence of wave effects in the non-linear regime leads to enhanced suppression of the power spectrum when compared to simulations where only the initial conditions had been modified. This agrees with the findings of \textsc{AXIREPO} and \textsc{AX-GADGET} codes~\cite{May2023TheHalo,Nori2018Ax-Gadget}. In simulations with $f=1$ and $m\sim10^{-22}$ eV, it was found that the wave effects can cause an increase in power on scales approaching $k\sim 1000\;h/$Mpc~\cite{May2023TheHalo} which we do not resolve sufficiently well with a $512^3$ grid (see Appendix~\ref{app:resolution}). In the low fraction limit, we found that the suppression of the power spectrum was mostly attributable to the change in the initial conditions rather than the wave effects. We also observed that the excess suppression due to the wave effects scales roughly linearly with $f$ if $f\lesssim 0.1$.

Next, we investigated the density profiles of halos with mixed dark matter. We found that, for high fractions $f\geq 0.50$, the halo exhibited a cored density profile matching closely the fitting formula of Ref.~\cite{Schive2014CosmicStructure} even in the presence of CDM. For a fraction of $f=0.5$, the CDM created a steep potential well exerting a pressure on the FDM soliton and caused its core radius to shrink. 
For lower fractions, we found that no stable core could form due to the enhanced gravitational potential in the presence of CDM. We compared the predicted size of the core (accounting for the CDM density) with the measured density profile of the FDM and concluded that the ground-state solution would have a radius large enough to be detected given our numerical resolution. Our findings that the soliton core does not form at a fraction of $f=0.1$ are in agreement with the results of spherical collapse simulations~\cite{Schwabe2020SimulatingMixed}.

We ran a customized halo finder and measured the halo mass function for pure CDM and 10\% FDM with a mass of $2.5\times10^{-22}$ eV. Our implementation of the halo finder is a combination of the friends-of-friends algorithm to identify the halo centers with CDM particles and the spherical overdensity finder to correct for the FDM density on a grid. The final halo finder gave us a halo catalog with mass measurements accounting for both dark matter components. We could not be conclusive in our \textsc{axionHMCode} comparison, but found a qualitative agreement given the limited statistics.
We also identified a low mass halo below the predicted cut-off introduced in the biased tracer halo model of Ref.~\cite{Vogt2023ImprovedMixed} and found that it had a very small FDM fraction, as expected.

As is the case for the power spectra, testing non-linear halo models further will benefit from cosmological simulations with a larger halo population and a lower final redshift. The time-step requirements for the Schr\"odinger-Poisson system make this task challenging and such simulations may require other computational approaches to evolve the FDM dynamics. This is exacerbated in the presence of a steep external gravitational potential created by the CDM which shrinks the de Broglie wavelength of the FDM thus increasing the resolution needed to study the halo cores.
We leave more involved analyses of halo statistics in mixed dark matter for future work.

Finally, we investigated the possibility that FDM could reduce the halo ellipticity around the soliton cores since a similar effect has been shown to occur for models of self-interacting dark matter. We expected the wave effects and modified initial conditions to isotropize the halo density on scales comparable to the FDM de Broglie wavelength and found this to be the case for the high FDM fraction ($f=0.50$). However, we found no trace of this phenomenon in halos with $f\leq 0.10$. It is possible that the impact of the wave effects manifested on scales which escaped the simulation resolution. In a mixed dark matter cosmology, we found that the FDM is not distributed equally across halos of different masses as displayed in Fig.~\ref{fig:half-half}.

In this study, we have run the first cosmological simulations of a mixture of cold and fuzzy dark matter. We have found that the resulting dark matter model combined features from both its constituents but exhibited a unique behaviour amongst known dark matter models. These findings will have a profound impact in the study of ultralight particles and the modelling of their behaviour on non-linear scales.

\begin{acknowledgments}
AL acknowledges support from NASA grant 21-ATP21-0145. AL would like to thank Jens Niemeyer, Benedikt Eggemeier, Mateja Gosenca, Tibor Dome, Simon May, Mathew Madhavacheril, J. Richard Bond, and Daniel Grin for useful discussions. RH acknowledge support from the Natural Sciences and Engineering Research Council of Canada. The Dunlap Institute is funded through an endowment established by the David Dunlap family and the University of Toronto. The authors at the University of Toronto acknowledge that the land on which the University of Toronto is built is the traditional territory of the Haudenosaunee, and most recently, the territory of the Mississaugas of the New Credit First Nation. They are grateful to have the opportunity to work in the community, on this territory. R. H. additionally acknowledges support from CIFAR, and the Azrieli and Alfred. P. Sloan Foundations. DJEM is supported by an Ernest Rutherford Fellowship from the STFC, Grant No. ST/T004037/1 and by a Leverhulme Trust Research Project (RPG-2022-145). Computations described in this work were performed with resources provided by the North-German Supercomputing Alliance (HLRN). The analysis was performed on the Niagara supercomputer at the SciNet HPC Consortium. SciNet is funded by Innovation, Science and Economic Development Canada; the Digital Research Alliance of Canada; the Ontario Research Fund: Research Excellence; and the University of Toronto~\cite{Scinet,Niagara}.
\end{acknowledgments}

\appendix
\section{Pade Approach to SP System}\label{app:pade}
We propose an alternative method for solving the system of equations~(\ref{eq:shooting}). The system is a boundary value problem with conditions at $r\to \infty$ and is therefore potentially unstable numerically. Moreover, it allows an infinite number of solutions, only one of which (the zero node solution) is stable~\cite{Marsh2015AxionDark}. The usual method is to first transform the system in dimensionless variables
\begin{align}
    r &\to \frac{m}{\hbar}r,\\
    \phi &\to \frac{\hbar\sqrt{4\pi G}}{m^2} \phi,\\
    \rho_\mathrm{CDM} &\to \frac{ 4\pi G \hbar^2}{m^2} \rho_\mathrm{CDM},\\
    \gamma &\to \frac{\hbar}{m}\gamma.
\end{align}
Then, we can use the scaling of the system using a scalar $\lambda$~\cite{Guzman2004EvolutionOf,Marsh2015AxionDark}
$\{r, \phi, \rho_\mathrm{CDM}, \gamma\} \to \{ \lambda^{-1} \tilde{r}, \lambda^{2} \tilde{\phi}, \lambda^{4} \tilde{\rho}_\mathrm{CDM}, \lambda^{2} \tilde{\gamma}\}$
where we choose $\lambda = \phi(r=0)^{1/2}$ so that $\tilde{\phi}(r=0)=1$. Our new rescaled system of equations is then
\begin{align}
    \frac{\partial^2 (\tilde r\tilde\phi)}{\partial \tilde r^2} &= 2 \tilde r\left(\tilde V- \tilde\gamma\right) \tilde\phi,\label{eq:rescaled-system}
    \\
     \frac{\partial^2 (\tilde r \tilde V)}{\partial \tilde r^2} &=\tilde r\left(\tilde\phi^2+\tilde \rho_\mathrm{CDM}\right),
\end{align}
with boundary conditions
\begin{align}
    \tilde\phi(0) &= 1,\\
    \tilde\phi(\tilde r\to\infty) &= 0\\
    \tilde\phi^\prime(0) &= 0,\\
    \tilde V^\prime(0) &=0\\
    \tilde V(r\to \infty) &=0,
\end{align}
where the prime denotes differentiation with respect to $r$. 

Some tricks have been suggested to approximate the problematic boundary conditions at infinity~\cite{Marsh2015AxionDark}, but here we consider a complementary approach to this problem which satisfies the exact boundary conditions and closely approximates $\phi$. Namely, we define the inverse polynomial estimator
\begin{align}
    \hat \phi(x) = \left(1+\sum_{k=1}^n a_k x^k\right)^{-1}.
\end{align}
This resembles the Pad\'e approximant with the condition that the polynomial in the numerator is set to unity. This ensures that the condition at $r=0$ is met. We note that the boundary conditions at infinity are trivially satisfied given that the inverse polynomial is monotonically decreasing. Given the success of the soliton fit with $\phi\propto r^{-8}$ at large $r$, we pick $n=8$.

\begin{figure}[t]
    \centering
    \includegraphics[width=\linewidth]{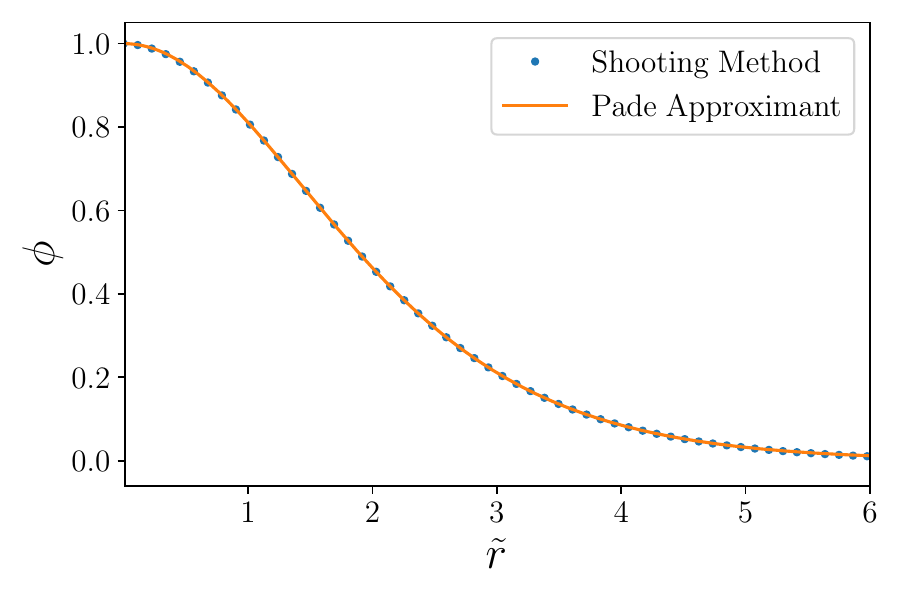}
    \caption{Comparison between the numerical solution using the shooting method and the Pad\'e approximant approach in the pure FDM case. We note an agreement to about $0.5\%$.}
    \label{fig:phi_comparison}
\end{figure}

From Ref.~\cite{Guzman2004EvolutionOf}, we can solve for the potential through
\begin{align}
    V(r)= V_0 + \int_0^r y \left(\phi^2+\rho_\mathrm{CDM}\right) dy -\frac{M(r)}{r},
\end{align}
where $M$ is the number density
\begin{align}
    M(r)\equiv  \int_0^r y^2 \left(\phi^2+\rho_\mathrm{CDM}\right) dy.
\end{align}
For simplicity, let us consider only the case where the CDM is absent. Then the potential $\tilde V^\prime \propto r^{1-N}$ for large $r$ and is equal to zero at $r=0$. The same goes for $\tilde \phi$. Finally, the boundary condition $\tilde V(r\to \infty)=0$ can be satisfied by an appropriate choice of $V_0$. The problem is now to find the coefficients $a_k$ so that $\hat \phi \approx \tilde\phi$ without having an analytic representation for $\tilde\phi$. For this, we borrow the definition of the \textit{loss function} implemented in physics-informed neural networks~\cite{Raissi2017PhysicsInformed}. We define
\begin{align}
    \mathcal{L} \equiv\frac{1}{N} \sum_{r_i \in R} \left[\partial^2_{\tilde r}\left(\tilde r_i \hat\phi\right)\bigg|_{r=r_i}- 2\tilde r_i  \left(\tilde V_i- \tilde\gamma\right) \hat\phi_i \right]^2,
\end{align}
where $\hat \phi_i = \hat \phi(r_i)$, $\tilde V_i = \tilde V(r_i)$, and where $R$ is a set of $N$ sample points between 0 and $r_{\rm max}$. We note here that we can be quite flexible in our choice of $r_{\rm max}$ given that our sample solution satisfies all the boundary conditions. To approximately solve this system, we then minimize the loss function over our coefficients $a_k$. We also have two other unknowns in $\tilde V_0$ and $\tilde \gamma$. However, given the form of Eq.~(\ref{eq:rescaled-system}), it is immediate that there is a degeneracy between the two parameters and that we can simply combine them in one $\tilde \gamma_0\equiv\tilde V_0-\tilde\gamma$. This gives a total $N+1$ free parameters over which to minimize. We employ the \textsc{basinhopping} method implemented with the \textsc{SciPy} Python package~\cite{Virtanen2020SciPy} for this procedure. Having found the $a_k$ coefficient, we compare our approximated solution to the numerical result when using the shooting method. The two solutions are plotted in Fig.~\ref{fig:phi_comparison} where we note the remarkable agreement between the two. The advantage of the method we develop here is that it allows us to solve the system for a variety of gravitational potential shapes without having to venture a guess about the value of $\gamma$. In other instances, this could be problematic as the wrong guess would lead to unstable solutions with a non-zero number of nodes.

\section{Impact of Box Length and Resolution} \label{app:resolution}
In Fig.~\ref{fig:nlmps_mass}, we observe a change in the wavenumber at which the matter power spectrum deviates form the linear theory prediction as a function of FDM mass. This is particularly visible for high masses $m\sim 10^{-21}$ eV. However, we expect to recover the CDM non-linear power spectrum in the limit of $m\to \infty$. The main difference in configuration between the simulations of different FDM masses is the choice of box length as listed in Table~\ref{tab:sim_list}. We investigate if this change in box size is responsible for the varying non-linear scales.

\begin{figure}
    \centering
    \includegraphics[width=\linewidth]{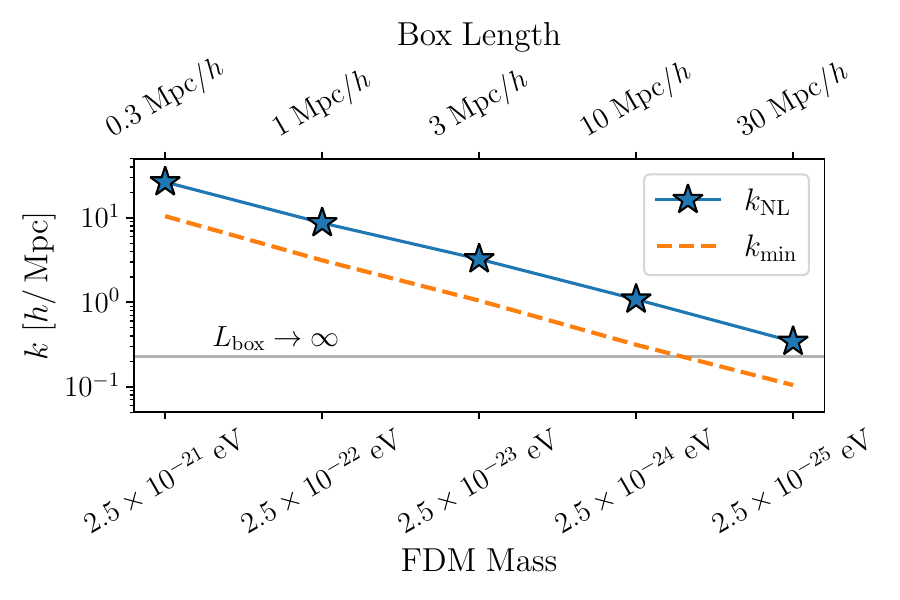}
    \caption{Non-linear and fundamental (minimum) scales for simulation boxes 3, 6, 7, 8, 9 (see Table~\ref{tab:sim_list}) as a function of FDM mass and box size. The grey line denotes the non-linear scale in the limit of an infinitely large simulation volume, matching the theoretical calculation. We observe a strong dependence of the non-linear scale on box size which matches the result of Fig.~\ref{fig:nlmps_mass} for $z=4$.}
    \label{fig:knl}
\end{figure}

First, we define what we mean by the non-linear scale which we label $k_{\rm NL}$. For this we use he approach of Ref.~\cite{Blas2013OnTheNon-Linear} and define
\begin{align}
    \sigma^2_d(k, k_{\rm min}) \equiv \frac{4\pi}{3}\int_{k_{\rm min}}^k dq\; P(q),
\end{align}
where $P$ is the linear matter power spectrum. From the above definition, we consider that non-linearities arise when $k\sigma_d \gg 1$.  In analytic calculations, we omit the lower bound on the integral since we can generate the linear matter power spectrum to very small wavenumbers. In our case, since some of our simulation volumes are very small, the coupling between large and small-scale modes is limited to the scales below the fundamental scale of the box. To account for this, we solve for the non-linear scale using
\begin{align}
    k_{\rm NL}\sigma_d(k_{\rm NL}, 2\pi/L_{\rm box}) = 5,\label{eq:knl-def}
\end{align}
where $L_{\rm box}$ is the size of the box. Note that the factor of five on the RHS of Eq.~(\ref{eq:knl-def}) is arbitrary, but our conclusion are largely unchanged for any number greater than unity but less than order 10. We display the calculated $k_{\rm NL}$ as a function of FDM mass and box size in Fig.~\ref{fig:knl}. All the calculations involved in this appendix are done assuming a redshift $z=4$.

\begin{figure}
    \centering
    \includegraphics[width=\linewidth]{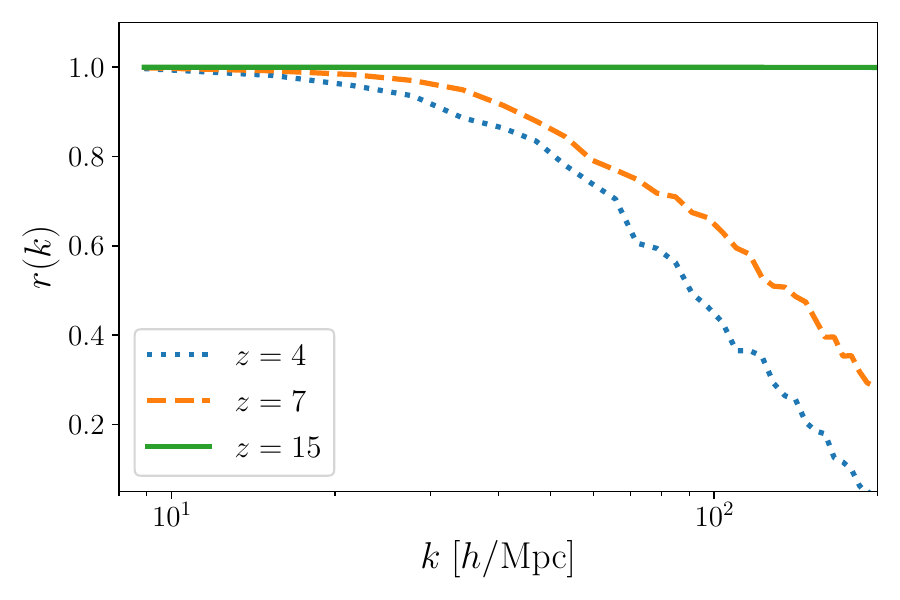}
    \caption{Cross-correlation coefficient of the $1024^3$ and $512^3$ resolution boxes as a function of redshift.}
    \label{fig:scaling-test}
\end{figure}

We also compare the density fields of boxes 3 and 10 which have the same cosmology, but different resolutions. We take the Fourier transform of the (total) density fields in both boxes at redshifts $z=15,\;7,\;4$ and compute the correlation coefficient
\begin{align}
    r(k)\equiv \frac{\langle\delta_{\rm LR}(\mathbf{k}) \delta_{\rm HR}(\mathbf{k}) \rangle}{\sqrt{\langle\delta_{\rm LR}^2(\mathbf{k}) \rangle \langle \delta_{\rm HR}^2 (\mathbf{k})\rangle}},
\end{align}
where $\delta_{\rm LR,\;HR}(\mathbf{k})$ are the density fields of the low and high-resolution boxes (respectively boxes 3 and 10). We plot the results of this calculation in Fig.~\ref{fig:scaling-test}.
At early times, the system is still very linear and the difference in resolution does not impact the density field. However, we note a degradation of the correlation coefficient at scales above $\sim 100\;h/$Mpc which we attribute to small-scale wave fluctuations around the mean density.

\bibliography{apssamp}% Produces the bibliography via BibTeX.

\end{document}